# An Introduction on Dependency Between Hardware Life Time Components and Dynamic Voltage Scaling

*Nasrin Jaberi*

Department of IT and Computer Engineering
Najafabad Payame Noor University
*Nasrinjaberi60@yahoo.com*

## 1. Introduction and motivation

The heart of a cloud computing infrastructure is the computing components which vary from a cluster to a supercomputer depend on the size of the cloud. As all processing inside a cloud is done in these components, they are the most energy consume parts of a cloud. Earth Simulator and Petaflop are two computing systems with 12 and 100 megawatts of peak power, respectively[1-2]. With an approximate price of $100 \frac{dollors}{megawatt}$, their energy costs during peak operation times are 1,200 and 10,000 dollars per hour; this is beyond the acceptable budget of many (potential) HPCS operators. In addition to power cost, cooling is another issue in a cloud infrastructure that must be addressed due to negative effects of high temperature on electronic components. The rising temperature of a circuit not only derails the circuit from its normal range but also results in decreasing the lifetime of its components. A formula based on Arrhenius Law indicates that the life expectancy of components decreases 50% for every $10^o C$ increase. In the opposite direction, the lifetime of components will be doubled for each $10^o C$ decrease[1, 3].

To reduce energy consumption, various issues such as resource management in both software and hardware must be addressed. Dynamic voltage-frequency scaling (DVFS) is an efficient energy saving method in designing new processors. Energy savings in this method is achieved by stretching tasks along processors' slack times. This is based on the fact that the power consumption in CMOS circuits has direct relation with frequency and the square of voltage supply. In this case, the execution time and power consumption are controllable by switching between processor's frequencies and voltages.

If a task has a deadline for completion, DVFS suggests stretching processing time of the task and reducing energy by decreasing processors' operating frequencies. A typical DVFS-enabled processor can operate in a number of different frequencies in its active mode. For example, AMD Turion MT-34 can operate at six frequencies ranging from 800MHz to 1800MHz [2, 4]. The power consumption of a processor consists of two parts: (1) dynamic part that is mainly related to CMOS circuit switching energy, and (2) static part that addresses the CMOS circuit leakage power. While a DVFS-enabled processor shows different power consumptions in different frequencies, other components of a system (such as Memory, Flash Drive, Bus, etc.) operate at a single frequency and spend the same power in both task's active and idle time[4]. For every task $Ti$, the power consumption is formulated as[5]:

$$\begin{cases} P_{Active,f,T_i} = A * f_{T_i} * V_{dd,T_i}^2 + B * V_{dd,T_i} + P_{Device} \\ P_{Idle,T_i} \simeq P_{Idle} \\ Energy_{T_i} = P_{Active,T_i} * t_{Active,f,T_i} + P_{Idle} * t_{Idle,f,T_i} \end{cases}$$

## 2. The open question

The main open question is how to calculate the effect of switching between frequencies in DVFS technique on the lifetime of the cluster components. As moving from one frequency to another in DVFS technique always gives a shock to the component and consequently decreases the component lifetime, therefore, it becomes interesting to answer the question of how fast a component can change its speed in order to decrease power without changing its lifetime. Our primary study shows that changing speed from one frequency to another has a direct relation with lifetime:

*lifetime* $\propto \Delta f$

Moreover, energy has a complex relation with the speed of frequency changes. Response to this problem, we propose a step-based frequency changes (Figure.1). As a processor uses a few numbers of frequencies ( $f_1 < f_2 < ... < f_N$ ), therefore moving from, for example, $f_1$ to $f_N$ can be divided into several steps as:

$$\begin{cases} f_1 \to f_2 \\ f_2 \to f_3 \\ \vdots \\ f_{N-1} \to f_N \end{cases}$$

We expect this step-based frequency changing preserves the components from harmful shocks (and therefore increase the lifetime).

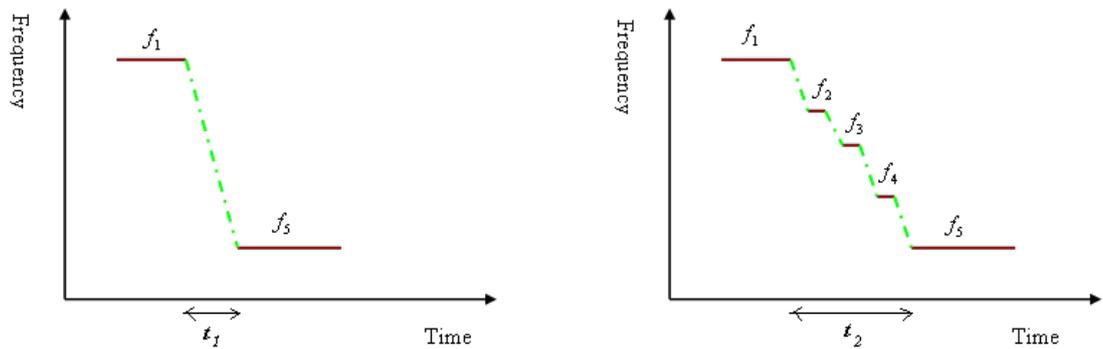

*Figure.1 Step-based frequency changes*

**References**

[1] R. Ge, *et al.*, "Performance-constrained Distributed DVS Scheduling for Scientific Applications on Power-aware Clusters," presented at the Proceedings of the 2005 ACM/IEEE conference on Supercomputing, 2005.

[2] N. B. Rizvandi, *et al.*, "Linear Combinations of DVFS-enabled Processor Frequencies to Modify the Energy-Aware Scheduling Algorithms," presented at the Proceedings of the 2010 10th IEEE/ACM International Symposium on Cluster, Cloud and Grid Computing (CCGrid), Melbourne, Australia, May 17-20, 2010.

[3] N. B. Rizvandi, *et al.*, "Multiple Frequency Selection in DVFS-Enabled Processors to Minimize Energy Consumption," in *Energy Efficient Distributed Computing Systems*, ed: Wiley, 2011.

[4] J. Zhuo and C. Chakrabarti, "Energy-efficient dynamic task scheduling algorithms for DVS systems," *ACM Trans. Embed. Comput. Syst.,* vol. 7, pp. 1-25, 2008.

[5] N. B. Rizvandi, *et al.*, "Some observations on optimal frequency selection in DVFS-based energy consumption minimization," *J. Parallel Distrib. Comput.,* vol. 71, pp. 1154-1164, 2011.